\documentstyle[preprint,aps]{revtex}

\begin{document}

\draft

\title{Spinor and Isospinor Structure of Relativistic
Particle Propagators}

\author{Dmitri M. Gitman and Sh.M.Shvartsman}

\address{Instituto de F\'{\i}sica, Universidade de S\~ao Paulo \\
Caixa Postal 20516-CEP 01498-970-S\~ao Paulo, S.P., Brasil \\
Department of Physics, Case Western Reserve University \\
Cleveland, OH 44106, USA
}

\date{\today}

\maketitle

\begin{abstract}

Representations by means of path integrals are used to find spinor and
isospinor structure of relativistic particle propagators in external fields.
For Dirac propagator in an external electromagnetic field all
grassmannian integrations are performed and a general result is
presented via a bosonic path integral. The spinor structure
of the integrand is given explicitly by its decomposition in the independent
$\gamma$-matrix structures. Similar technique is used to get the isospinor
structure of the scalar particle propagator in an external non-Abelian field.

\end{abstract}
\pacs{}

Propagators of relativistic particles in external fields
(electromagnetic, non-Abelian or
gravitational) contain important information about quantum behavior of
these particles. Moreover, if such propagators are
known in arbitrary external field, one can find exact
one-particle Green's functions in the corresponding quantum field theory,
taking functional integrals over the external field. Dirac propagator
in an external electromagnetic field distinguishes from one of a scalar
particle by a complicated spinor structure, and its explicit form was unknown
in general case until present time.  This
problem attracted attention of researchers already for a long time.
Feynman, who had written first a path integral for probability amplitude in
nonrelativistic quantum mechanics  \cite{F1} and then a path integral for
a scalar particle propagator
\cite{F2}, had also attempted to derive a representation for Dirac
propagator via a bosonic path integral \cite{F3}.
After Berezin had introduced the integral over grassmannian variables, it
turned out to be
natural to present Dirac propagator via both bosonic and grassmannian path
integrals. Such representations have been
discussed in the literature for a long time in different contexts
\cite{b1,b2,b3,b4,b5,b6,b7,b8,b9,b10,b11}. Nevertheless,
attempts  to write Dirac propagator via a
bosonic path integral only where continued. So, Polyakov \cite{b12} assumed
that the propagator
of free Dirac electron in $D=3$ Euclidean space-time can be presented by
means of a bosonic path integral, similar to a scalar particle, modified by
a so called spinor factor. This idea was developed in \cite{b13} to write
spinor factor for Dirac fermions, interacting with a non-Abelian gauge
field in $D$ dimensional Euclidean space-time. In that representation the
spinor factor itself was presented via some additional bosonic path integrals.
One ought to say that sometimes it is possible to find
Dirac propagator in special configurations of electromagnetic
field, thus, in these particular cases its spinor structure can be described
explicitly. In fact, there are only few configurations of external field, where
that can be done: constant homogeneous electromagnetic field \cite{b14},
electromagnetic plane wave \cite{b15,b2}, crossed electric and magnetic
fields \cite{b16,b17}, and combination of constant homogeneous electromagnetic
field with plane wave field \cite{b3,b18,b9}.

In this paper we consider a representation of Dirac propagator in arbitrary
electromagnetic field as a path
integral over bosonic and grassmannian variables and demonstrate that all
grassmannian integrations can be performed so that a result can only be
presented via a bosonic path integral over coordinates; the integrand of this
path integral differs from the corresponding expression in scalar case by a
spin factor, which  spinor structure
is given explicitly. Similar technique can be used to get
the isospinor structure of the scalar particle propagator in an external
non-Abelian field.

The propagator of a spinning particle in an external electromagnetic
field $A_{\mu}(x)$ is the causal Green's function $S^{c}(x,y)$ of Dirac
equation in this field,
\begin{equation}
\label{e1}
\left[\gamma^{\mu}\left(i\partial_{\mu}-gA_{\mu}(x)\right)-m\right]S^{c}(x,y)
=-\delta^{4}(x-y)\;,\;
\end{equation}
where $x=(x^{\mu}),\;\left[\gamma^{\mu}\;,\;\gamma^{\nu}\right]_{+}
=2\eta^{\mu \nu},\;\eta^{\mu \nu}={\rm diag}(1-1-1-1),\;
\mu,\nu=\overline{0,3}$.

Consider a lagrangian form of the path integral
representation (see \cite{b10}) for, transformed by $\gamma^{5}=
\gamma^{0}\gamma^{1}\gamma^{2}\gamma^{3}$  function  $\tilde{S}^{c}(x,y)=
S^{c}(x,y)\gamma^{5}$,
\begin{eqnarray}\label{e2}
&&\tilde{S}^{c}=\tilde{S}^{c}(x_{out},x_{in})=\exp\left\{i\gamma^{n}
\frac{\partial_{\ell}}{\partial \theta^{n}}\right\}\int_{0}^{\infty}de_{0}
\int d\chi_{0}\int_{e_{0}}
De\int_{\chi_{0}}D\chi \int_{x_{in}}^{x_{out}}Dx \nonumber \\
&&\times\int D\pi_{e}\int D\pi_{\chi} \int_{\psi(0)+\psi(1)=\theta}
{\cal D}\psi M(e)
\exp\left\{i\int_{0}^{1}\left[-\frac{\dot{x}^{2}}
{2e}-\frac{e}{2}m^{2}-g\dot{x}A(x)+iegF_{\mu \nu}(x)\psi^{\mu}\psi^{\nu}\right.
\right.\nonumber \\
&&\left.\left.\left. +i\left(\frac{\dot{x}_{\mu}\psi^{\mu}}{e}-m\psi^{5}\right)
\chi-i\psi_{n}\dot{\psi}^{n}+\pi_{e} \dot{e}+\pi_{\chi} \dot{\chi}\right]d\tau
+ \psi_{n}(1)\psi^{n}(0)\right\}\right|_{\theta=0}\;,
\end{eqnarray}
\noindent where
$\left[\gamma^{m}\;,\;\gamma^{n}\right]_{+}=2\eta^{mn}, \; m,n=\overline
{0,3},5, \; \;
 \eta^{mn}={\rm diag}(1-1-1-1-1)$;
\noindent $\theta^{n}$ are auxiliary grassmannian (odd) variables,
anticommuting by definition with the $\gamma$-matrices; $x^{\mu}(\tau),
\; e(\tau), \; \pi_{e}(\tau)$ are bosonic trajectories of
integration; $\psi^{n}(\tau), \; \chi(\tau), \; \pi_{\chi}(\tau)$ are odd
trajectories of integration; and boundary conditions
$x(0)=x_{in}, \;x(1)=x_{out},\;
e(0)=e_{0},\;\psi^{n}(0)+\psi^{n}(1)=\theta^{n}, \;
\chi(0)=\chi_{0} $ take place;
\begin{equation}\label{e3}
M(e)=\int Dp\exp\left\{ \frac{i}{2}\int_{0}^{1} ep^{2}d\tau\right\}, \;
{\cal D}\psi=D\psi\left[\int_{\psi(0)+\psi(1)=0} D\psi \exp\left\{\int_{0}^{1}
 \psi_{n}\dot{\psi}^{n}
d\tau \right\}\right]^{-1} \; .
\end{equation}

We are going to demonstrate that the propagator (\ref{e2}) can
be only expressed  through a bosonic path integral over coordinates $x$. To
this end one needs to fulfil several functional
integrations. First, one can integrate over $\pi_{e}$ and $\pi_{\chi}$,
and then use arisen $\delta$-functions to remove the functional integration
over $e$ and
$\chi$,
\begin{eqnarray*}
&&\tilde{S}^{c}=-\exp\left\{i\gamma^{n}
\frac{\partial_{\ell}}{\partial \theta^{n}}\right\}
\int_{0}^{\infty}de_{0}\;\int_{x_{in}}^{x_{out}}Dx\;
\int_{\psi(0)+\psi(1)=\theta} {\cal D}\psi\;M(e_{0})
\;\int_{0}^{1} \left(\frac{\dot{x}_{\mu}\psi^{\mu}}{e_{0}}-
m\psi^{5}\right)d\tau \\
&&\left.\times \exp\left\{i\int_{0}^{1}\left[-\frac{\dot{x}^{2}}{2e_{0}}-
\frac{e_{0}}{2}m^{2}-g\dot{x}A(x)+ige_{0}F_{\mu \nu}(x)\psi^{\mu}\psi^{\nu}-
i\psi_{n}\dot{\psi}^{n}\right]d\tau+ \psi_{n}(1)\psi^{n}(0)\right\}
\right|_{\theta=0}\;.
\end{eqnarray*}
Then, it is convenient to replace the integration over $\psi$ by one
over related odd velocities $\omega$,
\begin{equation}\label{e5}
\psi(\tau)=\frac{1}{2}\int_{0}^{1} \varepsilon (\tau-\tau')\omega(\tau')d\tau'+
\frac{1}{2}\theta\;,\;\;\omega(\tau)=\dot{\psi}(\tau)\;,\;
\varepsilon (\tau)={\rm sign}\;\tau\;.
\end{equation}
There are not more any restrictions on $\omega$; because of
(\ref{e5}) the boundary conditions for $\psi$ are obeyed automatically.
The corresponding
Jacobian does not depend on variables and cancels with the same one from the
measure (\ref{e3}). Thus\footnote{Here and further,
we are using condensed notations, e.g. $\omega\varepsilon
\omega=\int_{0}^{1} d\tau d\tau'\omega(\tau)\varepsilon(\tau-\tau')\
\omega(\tau')$ and so on.} ,
\begin{eqnarray*}
&&\tilde{S}^{c}=-\frac{1}{2}\exp\left\{i\gamma^{n}
\frac{\partial_{\ell}}{\partial \theta^{n}}\right\}
\int_{0}^{\infty}de_{0}\;\int_{x_{in}}^{x_{out}}Dx\;
\int {\cal D}\omega\; M(e_{0})\left[\frac{\dot{x}_{\mu}}{e_{0}}\left(
\varepsilon \omega^{\mu}+\theta^{\mu}\right)-
m\left(\varepsilon \omega^{5}+\theta^{5}\right)\right]  \\
&&\times\exp\left\{i\left[-\frac{\dot{x}^{2}}{2e_{0}}-\frac{e_{0}}{2}
m^{2}-g\dot{x}A(x)\left.-\frac{ie_{0}g}{4}\left(\omega^{\mu}\varepsilon-
\theta^{\mu}\right)F_{\mu \nu}(x)\left(\varepsilon \omega^{\nu}+
\theta^{\nu}\right)+\frac{i}{2}\omega_{n}\varepsilon\omega^{n}
\right]\right\}\right|_{\theta=0}\;,
\end{eqnarray*}
\noindent where the measure ${\cal D}\omega$ is
\[
{\cal D}\omega=D\omega\left[\int D\omega \exp\left\{-\frac{1}{2}
\omega^{n}\varepsilon
\omega_{n}\right\}\right]^{-1}\;.
\]
\noindent One can prove, that for a function $f(\theta)$ in the Grassmann
algebra, the following identity holds
\begin{eqnarray}\label{e8}
&&\left.\exp\left\{i\gamma^{n}
\frac{\partial_{\ell}}{\partial \theta^{n}}\right\}f(\theta)\right|
_{\theta=0}=\left.f\left(\frac{\partial_{\ell}}
{\partial \zeta}\right)
\exp\left\{i\zeta_{n}\gamma^{n}\right\}\right|_{\zeta=0}
\nonumber \\
&&=\left.\sum_{k=0}^{4}\sum_{n_{1}\cdots n_{k}}f_{n_{1}\cdots n_{k}}
\frac{\partial_{\ell}}{\partial \zeta_{n_{1}}}\cdots
\frac{\partial_{\ell}}{\partial \zeta_{n_{k}}}
\sum_{l=0}^{4}\frac{i^{l}}{l!}\left(\zeta_{n}\gamma^{n}
\right)^{l}\right|_{\zeta=0}\; ,
\end{eqnarray}
where $\zeta_{n}$ are some odd
variables. Using (\ref{e8}), we get
\begin{eqnarray}\label{e9}
&&\tilde{S}^{c}=-\frac{1}{2}\int_{0}^{\infty}de_{0}\;\int_{x_{in}}^{x_{out}}
Dx\;M(e_{0})\left[\frac{\dot{x}_{\mu}}{e_{0}}\left(\varepsilon\frac{\delta_
{\ell}}{\delta \rho_{\mu}}+\frac{\partial_{\ell}}{\partial \zeta_{\mu}}\right)
-m\left(\varepsilon\frac{\delta}{\delta \rho_{5}}+i\gamma^{5}\right)\right]
\nonumber \\
&&\times\exp\left\{i \left[-\frac{\dot{x}^{2}}{2e_{0}}-\frac{e_{0}}{2}
m^2-g\dot{x}A(x)+\frac{ie_{0}g}{4}F_{\mu \nu}(x)
\frac{\partial_{\ell}}{\partial \zeta_{\mu}}
\frac{\partial_{\ell}}{\partial \zeta_{\nu}}\right] \right\} \nonumber \\
&&\left.\times R\left[x,\rho,\frac{\partial_{\ell}}{\partial \zeta}\right]
\exp\left\{i\zeta_{\mu}\gamma^{\mu}\right\}\right|_{\rho=0\;,\;\zeta=0}\;,
\end{eqnarray}
with
\begin{eqnarray}\label{e10}
&&R\left[x,\rho,\frac{\partial_{\ell}}{\partial \zeta}\right]
=\int {\cal D}\omega\exp\left\{-\frac{1}{2}
\omega^{n}T_{nk}(x|g)\omega^{k}+I_{n}\omega^{n}\right\}, \\
&&I_{\mu}=\rho_{\mu}-\frac{e_{0}g}{2}
\frac{\partial_{\ell}}{\partial \zeta_{\nu}}
F_{\nu \mu}(x)\varepsilon,\; I_{5}=\rho_{5}, \nonumber
\end{eqnarray}
\begin{equation}
\label{e11}
T_{nk}(x|g)=\left ( \begin{array}{cc}
          \Lambda_{\mu \nu}(x|g) & 0\\

          0                               &-\varepsilon\\
        \end{array}\right )
\;,\;\;\Lambda_{\mu \nu}(x|g)=\eta_{\mu \nu}\varepsilon
-\frac{e_{0}}{2} \varepsilon gF_{\mu \nu}(x)
\varepsilon\;.
\end{equation}
\noindent where $\rho_{n}(\tau)$
are  odd sources for $\omega^{n}(\tau)$. Integral in (\ref{e10})
is gaussian one. It can be easily done
\cite{b19}, remembering its original definition \cite{b10},
\begin{equation}
\label{e12}
R\left[x,\rho,\frac{\partial_{\ell}}{\partial \zeta}\right]=
\left[\frac{{\rm Det}T(x|g)}{{\rm Det}T(x|0)}\right]^{1/2}
\exp\left\{-\frac{1}{2}I_{n}
\left[T^{-1}(x|g)\right]^{nk}I_{k}\right\}\;,
\end{equation}
\noindent The ratio ${\rm Det}T(x|g)/{\rm Det}T(x|0)$ in (\ref{e12}) can be
replaced by ${\rm Det}\Lambda(x|g)/{\rm Det}\Lambda(x|0)$, due to the
structure (\ref{e11}) of the matrix $T(x|g)$, and the latter can be
presented in a convenient form, which allows one to avoid problems with
calculations of determinants of matrices with continuous indices,
\begin{equation}
\frac{{\rm Det}\Lambda(x|g)}{{\rm Det}\Lambda(x|0)}
=\exp\left\{-e_{0}\int_{0}^{g}dg'
\;{\rm Tr}\;{\cal G}(x|g') F(x)\right\},\;\;
{\cal G}^{\mu \nu}(x|g)=\frac{1}{2}\varepsilon
\left[\Lambda^{-1}(x|g)\right]^{\mu \nu}\varepsilon\;.
\label{e13}
\end{equation}
\noindent Substituting (\ref{e13}) into (\ref{e9}), and performing
functional differentiations with respect to $\rho_{\mu}$, we get
\begin{eqnarray}
\label{e14}
&&\tilde{S}^{c}=-\frac{1}{2}\int_{0}^{\infty}de_{0}\;\int_{x_{in}}^{x_{out}}
Dx\;M(e_{0})\left[\frac{\dot{x}^{\mu}}{e_{0}}K_{\mu \nu}(x)
\frac{\partial_{\ell}}{\partial \zeta_{\nu}}
-im\gamma^{5}\right]
\exp\left\{i \left[-\frac{\dot{x}^{2}}{2e_{0}}-\frac{e_{0}}{2}m^2 \right.
\right.\nonumber \\
&&\left.\left.\left.-g\dot{x}A(x)+\frac{ie_{0}}{2}\int_{0}^{g}dg'
\;{\rm Tr}\;{\cal G}(x|g') F(x)+\frac{ie_{0}g}{4}\left(F(x)
K(x)\right)_{\mu \nu}
\frac{\partial_{\ell}}{\partial \zeta_{\mu}}
\frac{\partial_{\ell}}{\partial \zeta_{\nu}}\right] \right\}
\exp\left\{i\zeta_{\mu}\gamma^{\mu}\right\}\right|_{\zeta=0}\;,
\nonumber\\
&&K_{\mu \nu}(x)=\eta_{\mu \nu}+e_{0}g\left({\cal G}(x|g)
F(x)\right)_{\mu \nu}\;.
\end{eqnarray}
The differentiation over $\zeta$ in (\ref{e14}) can be fulfilled
explicitly, using eq. (\ref{e8}). Thus, finally
\begin{eqnarray}
\tilde{S}^{c}&=&\frac{i}{2}\int_{0}^{\infty}de_{0}\;\int_{x_{in}}^{x_{out}}
Dx\;M(e_{0})\Phi(x,e_{0})
\exp\left\{i \left[-\frac{\dot{x}^{2}}{2e_{0}}-\frac{e_{0}}{2}
m^2 -g\dot{x}A(x)\right]\right\}\;,
\label{e15}\\
\Phi(x,e_{0})&=&\left[m+(2e_{0})^{-1}\dot{x}K(x)\left(1-2gF(x)K(x)\right)
\gamma+im\frac{e_{0}g}{4}\left(F(x)K(x)\right)_{\mu \nu}\sigma^{\mu \nu}
\right. \nonumber \\
&&\left.+i\frac{g}{4}\left(\dot{x}K(x)\gamma\right)\left(F(x)K(x)\right)
_{\mu \nu}\sigma^{\mu \nu}
+m\frac{e^{2}_{0}g^{2}}{16}\left(F(x)K(x)\right)^{\ast}_{\mu \nu}
\left(F(x)K(x)\right)^{\mu \nu}\gamma^{5}\right]
\nonumber \\
&&\times\exp\left\{-\frac{e_{0}}{2}\int_{0}^{g}dg'
\;{\rm Tr}\;{\cal G}(x|g') F(x)\right\}\;,
\label{e16}
\end{eqnarray}
where $\sigma^{\mu \nu}=\frac{i}{2}\left[\gamma^{\mu}\;,\;\gamma^{\nu}\right]
_{-}\;,\;\left(F(x)K(x)\right)^{\ast}_{ \mu \nu}
=\frac{1}{2}\epsilon_{\mu \nu
\alpha \beta}\left(F(x)K(x)\right)^{\alpha \beta}$, and $\epsilon_{\mu \nu
\alpha \beta}$ is Levi-Civita symbol.

The eq.(\ref{e15}) gives a representation for Dirac propagator
as a path integral over bosonic
trajectories of a functional, which spinor structure is found
explicitly, namely, its decomposition in all independent
$\gamma$-structures is given. The functional $\Phi(x,e_{0})$ can be called
spin factor, and namely it distinguishes Dirac propagator from the
scalar one. One needs to stress that spin factor is gauge invariant, because
of its dependence of $F_{\mu \nu}(x)$ only.

In the same manner one can describe the isospinor structure of relativistic
particle propagators. Here we restrict ourselves with a consideration of
a scalar particle propagator in an external electromagnetic $A_{\mu}(x)$
and non-Abelian $B_{\mu}(x)$ fields. Such a propagator is the causal
Green's function $D^{c}(x,y)$ of the Klein-Gordon equation in the fields,
\begin{equation}
\left[\left(i\partial-gA(x)-B^{a}(x)T_{a}\right)^{2}-m^{2}\right]
D^{c}(x,y)=-\delta^{4}(x-y)\;,
\label{e17}
\end{equation}
where $T^{a}$ are generators of a corresponding group. Choosing for
simplicity $SU(2)$ as the group, we have $T_{a}=\frac{1}{2}\sigma_{a}$,
where $\sigma_{a}$ are Pauli matrices.
The propagator $D^{c}$ can be presented via bosonic and grassmannian path
integrals \cite{b7,b9},
\begin{eqnarray}
&&D^{c}=D^{c}(x_{out},x_{in})=\frac{i}{2}\exp\left\{i\sigma_{a}
\frac{\partial_{\ell}}{\partial \theta_{a}}\right\}\int_{0}^{\infty}de_{0}
\int_{e_{0}}De \int_{x_{in}}^{x_{out}}Dx\int D\pi_{e}
\int_{\phi(0)+\phi(1)=\theta} {\cal D}\phi \nonumber \\
&&\times M(e)
\left.\exp\left\{i\left[-\frac{\dot{x}^{2}}
{2e}-\frac{e}{2}m^{2}-g\dot{x}A(x)-\dot{x}B^{a}(x){\cal T}_{a}
-i\phi_{a}\dot{\phi}_{a}+\pi_{e} \dot{e}\right]+\phi_{a}(1)\phi_{a}(0)\right\}
\right|_{\theta=0}\;,
\label{e18}\\
&&{\cal D}\phi=D\phi \left[\int_{\phi(0)+\phi(1)=0}
 D\phi \exp\left\{\phi_{a}\dot{\phi}_{a}\right\}
\right]^{-1}\;,
\nonumber
\end{eqnarray}
where  $\theta_{a}$ are auxiliary odd variables, anticommuting by definition
with the $\sigma$-matrices; $\phi_{a}(\tau)$ are odd trajectories of
integration and ${\cal T}_{a}=-i\epsilon_{abc}\phi_{b}\phi_{c}$.
All grassmannian integrals can be done
similar to the spinning particle case and final result presented in the form
\begin{eqnarray}
\label{e19}
D^{c}&=&\frac{i}{2}\int_{0}^{\infty}de_{0}\int_{x_{in}}^{x_{out}}Dx\;
M(e_{0})\;\Phi(x)
\exp\left\{i\left[-\frac{\dot{x}^{2}}{2e_{0}}-\frac{e_{0}}{2}m^{2}-
g\dot{x}A(x)\right]\right\}\;,\\
\Phi(x)&=&\left[1+{\rm Tr}\;R(x){\cal G}(x|1)R(x)\;-\frac{i}{2}
L_{ab}(x)\epsilon_{abc}T_{c}\right]
\exp\left\{-\frac{1}{2}\int_{0}^{1}d\lambda \;{\rm Tr}\;{\cal G}(x|\lambda)
R(x)\right\}\;,
\label{e20}\\
&&{\cal G}(x|\lambda)=\frac{1}{2}\varepsilon Q^{-1}(x|\lambda)\varepsilon\;,\;
Q(x|\lambda)=\varepsilon I-\frac{\lambda}{2}\varepsilon R(x)
\varepsilon\;,\;R_{ab}(x)=\dot{x}B^{c}(x)\epsilon_{cab}\;,
\nonumber\\
&&L_{ab}(x)=R_{ab}(x)-\left[R(x){\cal G}(x|1)R(x)\right]_{ab}\;,
\nonumber
\end{eqnarray}
where $I$ is unit matrix in the group space.
The isospinor factor (\ref{e20}) in (\ref{e19}) is
presented by its decomposition in the generators $T_{a}$ of the $SU(2)$ group.
Explicit description of the spinor and isospinor structure of Dirac propagator
in both Abelian and non-Abelian external fields is more complicated problem
which, nevertheless, can be solved in the frame of the same approach.

{\large{\bf Acknowledgment}}

The authors would like to thank Professor J. Frenkel for
discussions.

\end{document}